\documentstyle[12pt]{article}
\begin{document}
\textwidth 160mm
\textheight 240mm
\topmargin -20mm
\oddsidemargin 0pt
\evensidemargin 0pt
\newcommand{\beq}{\begin{equation}}
\newcommand{\eeq}{\end{equation}}
\begin{titlepage}
\begin{center}

{\bf On Infrared Universality in Massive Theories. Another Example.}

\vspace{1.5cm}

{\bf  K.G.Selivanov }

\vspace{1.0cm}

{ITEP,B.Cheremushkinskaya 25,Moscow,Russia}

\vspace{1.9cm}

{ITEP-16/96, hep-thxxxxxxx}

\vspace{1.0cm}

\end{center}

%\begin{document}
%\maketitle
%vspace{3.5 cm}

\begin{abstract}
The infrared behaviour of the ${\phi}^3$-theory is 
discussed stressing analogies with the Witten-Seiberg
story about $N=2$ $QCD$. Though the microscopic theory is apparently
not integrable, the effective theory is shown
to be integrable at classical level, and a general solution
of it in terms of hypergeometric functions is obtained.
An effective theory for the multiparticle soft scattering is sketched. 
\end{abstract}
%\vfill
\end{titlepage}

%\begin{document}
\newpage
\setcounter{equation}{0}

1. Recently, very remarkable progress in understanding quantum field
theory has been developed \cite{wise}. The authors managed to obtain
the exact low-energy wilsonian-type effective action for the  $4D$ $N=2$
supersymmetric $SU(2)$ $QCD$ (see refs. \cite{wsadepts} for other
gauge groups and for diffferent matter multiplets). In \cite{gurpit} it was
noticed that the effective action gets 
nicely interpreted in terms of a solution of the so called Whitham equations
for some particular $2D$ integrable system. It was a sort of phenomenological,
though quite exciting, observation. The idea of a ``Whitham 
universality'' was put forward there. In \cite{japan} a plenty of formulas and
references for the 
future study of the universality was accumulated. 
See, also, refs. \cite{gurpitadepts} for other explorations of the subject.  

Here I consider the infrared behaviour of the ${\phi}^3$ field theory,
which is simpler formulated (compared to $N=2$ $QCD$) but is more
generic type of quantum field theory (with no symmetries, supersymmetries,
integrabilities etc. in it). Nevertheless, many of the concepts
exploited in the Witten-Seiberg consideration - the elliptic curves and their
moduli space, the modular symmetry, part of which playing the role of
duality transformation between week and strong couplings,
the massless-state singularities in the moduli space,
the fuchsian equations, etc. - show up here, too.
 
The infrared limit of a massive quantum field theory is
understood in the sense, that all degrees of freedom with spatial momenta above
some scale (small compared to the mass scale of the theory) are integrated
out (the typical time-momenta (typical energies) are then close to the
mass scale). That is this limit which enjois the nice 
universal properties listed above.

Of course, in nonintegrable theory is impossible to exactly integrate out
the fast modes. However, it is possible to integrate out them in the
classical approximation (which, as explained below, is equivalent to the
Whitham averaging  \cite{whitham} of the microscopic equations of motion).
And once the effective slow-variables theory is obtained, the
decoupling theorem \cite{appcarr} garanties that 
quantum corrections to the integration over
the fast degrees of freedom reduce just to renormalization of coupling
constants in the effective theory (modulo power terms in the ratio of
the typical scales), provided the effective field theory is renormalizable
by itself.

Since the procedure descibed is a more or less straightforward treatment of 
the original theory it allows to relate operators of the microscopic theory
to those of the effective theory, thus giving one not only 
the effective action but also the relevant observables, which is new
compared to the Witten-Seiberg discussion.

The main motivation for the construction was to describe the soft
multiparticle scattering in the tree approximation - the problem,
intensively discussed in
literature over the last five years (see reviews \cite{volrub} and refs.
therein). Some elements of the papers \cite{volgor}, \cite{libtr} can be
identified as fragmments of the picture exposed here.
Some preliminaries for the current discussion were published in 
\cite{agks}. 

I would like to stress that the ${\phi}^3$ is taken just because the things
are most clear in this case. The method is applicable
to any massive  theory whose
equations of motion are integrable in the spatially uniform case.
With some modifications it is also applicable to massless theories.

2. So, I consider the theory with lagrangian

\beq
\label{lagran}
L=\frac{1}{2}{(\partial{\phi})}^2+\frac{m^2}{2}{\phi}^{2}
+\frac{1}{3}{\lambda}{\phi}^{3}
\eeq
and as a rule I asume that $m=1$ while ${\lambda}$ is small compared to $1$.
I would like to integrate out all the modes with spatial momenta
above some scale ${\epsilon}$ for ${\epsilon}$ small compared to $m$.
To do this in the tree approximation, in other words, on the level of 
equations of motions means to find classical solutions without fast space
dependence for any
given slow space dependence. Experts will
immediately recognize this as just a typical task for the
Whitham-averaging method (some russian physicists call it
the Kapitsa's method, some russian mathematicians call it the Bogolyubov's
one). Anyway, it works as follows.

Take the general solution of the equations of motion with no
spatial dependence at all. For the ${\phi}^3$-theory it is
\beq
\label{solution}
\phi(t)=-\frac{6}{\lambda}(\frac{1}{12}
+\wp(t-t_{0}|{\omega},{\omega}') 
\eeq
where $t$ is, of course, time and ${\wp}(u|{\omega},{\omega}')$ is the
elliptic Weierstrass function with semiperiods $\omega$ and ${\omega}'$.
\beq
\label{weier}
\wp(u|{\omega},{\omega}')=\frac{1}{u^2}
+{\sum}'\frac{1}{(u-2m\omega-2m'{\omega}')^{2}}
-\frac{1}{(2m\omega+2n{\omega}')^{2}}
\eeq
From the equation of motion, corresponding to the lagrangian
Eq.~(\ref{lagran}), follows that the standard $g_{2}$-parameter is equal to
$\frac{1}{12}$:
\beq
\label{gparameter}
g_{2}({\omega},{\omega}')=60{\sum}' \frac{1}{(2m\omega
 + 2n{\omega}')^{4}}=\frac{1}{12}
\eeq
thus one of the ${\omega}$'s is expressed in terms of the other,
and the solution from Eq.~(\ref{solution}) depends on just two parameters,
$t_{0}$ and, say, ${\tau}=\frac{{\omega}'}{\omega}$, as it must be
for a general solution of the second order differential equation.

(Notice that in massless theory the $g_{2}$ parameter is equal to zero,
and the parameter ${\tau}=\frac{{\omega}'}{\omega}$ is
$-\frac{1}{2}+i\frac{\sqrt{3}}{2}$), and again there are just two
parmeters).

Then the Whitham procedure provides one with ${\phi}(t,x)$ 
\beq
\label{solution2}
\phi(t,x)=-\frac{6}{\lambda}(\frac{1}{12}
+({\nu}^2-k^2){\wp}(u|{\pi},{\pi}{\tau})) 
\eeq
which is a solution of the original equation, provided $\tau$, $\nu=u_{t}$,
$k=u_{x}$
are slow functions of $(t,x)$ obeying the following set of equations:
\begin{eqnarray}
\label{kapitsa}
\partial_{t}[{\nu}({\nu}^2-k^2)^{2}\int_{0}^{2{\pi}}du 
(\partial_{u}{\wp}(u|{\pi},{\pi}{\tau}))^{2}]=\nonumber\\
\partial_{x}[k({\nu}^2-k^2)^{2}\int_{0}^{2{\pi}}du 
(\partial_{u}{\wp}(u|{\pi},{\pi}{\tau}))^{2}]
\end{eqnarray}

\beq
\label{integrability}
\partial_{x}{\nu}=\partial_{t}k
\eeq

\beq
\label{mass}
\frac{12}{(2{\pi})^4}g_{2}({\tau}){\rho}^{4}=1
\eeq
where ${\rho}=\sqrt{{\nu}^2-k^2}$.
Some comments are in order here. The periods of the Weierstrass function in
Eq.~(\ref{solution2}) are rescaled so that the solution 
is $2{\pi}$-periodic in $u$ and the $u$-integration in Eq.~(\ref{kapitsa})
(the`` Whitham averaging'') is taken over the period. The Eq.~(\ref{mass})
is actually a consequence of Eq.~(\ref{gparameter}), $g_{2}(\tau)$ is
$g_{2}({\omega},{\omega}')$ with $2{\omega}=1$ and $2{\omega}'={\tau}$.
The Eq.~(\ref{mass}) expresses $\tau$ in terms of $({\nu},k)$ thus closing
the set of equations. Note that dimensionality of space doesn't matter here;
it is obvious how to insert space indexes in the equations above.

It can be shown that
\beq
\label{pert}
\frac{12}{(2{\pi})^{2}}g_{2}(\tau)=1+240\sum_{m=1}\frac{m^{3}q^{2m}}
{(1-q^{2m})}
\eeq
where $q={\exp}(i{\pi}{\tau})$, and also that
\beq
\label{integral}
\int_{0}^{2{\pi}}du 
(\partial_{u}{\wp}(u|{\pi},{\pi}{\tau}))^{2}=
\sum_{m=1}\frac{m^{4}q^{2m}}
{(1-q^{2m})^{2}}=\frac{1}{480i{\pi}}\partial_{\tau}g_{2}({\tau})
\eeq
These formulas are useful in the perturbative
treatment of Eq.~(\ref{kapitsa}) since, as will be clear later, $q$
in perturbative region is a small parameter. Using them the
Eq.~(\ref{kapitsa}) can rewritten as
\beq
\label{kap}
\partial_{t}[\frac{\nu}{\rho}(\partial_{\tau}{\rho})]= 
\partial_{x}[\frac{k}{\rho}(\partial_{\tau}{\rho})]
\eeq   

The Eqs.~(\ref{kap}),(\ref{integrability}),(\ref{mass}) are equations
of motion
of the infrared effective field theory, the corresponding lagrangian reading
\beq
\label{infrared}
L_{eff}(\tau)=(\frac{6}{\lambda})^{2}[\frac{1}{20i{\pi}}
\frac{\partial_{\tau}g_{2}({\tau})}{g_{2}(\tau)^{\frac{3}{2}}}
+\frac{1}{(2\pi)^{2}}\frac{g_{3}(\tau)}{g_{2}(\tau)^{\frac{3}{2}}}]
\eeq
One more personage of the theory of elliptic functions, 
the $g_{3}$-parameter, has entered the Eq.~(\ref{infrared}),
\beq
\label{gthree}
g_{3}(\tau)=140{\sum}' \frac{1}{(m+n{\tau})^{6}}
\eeq
$\tau$ in Eq.~(\ref{infrared}) is related to the
$\rho=\sqrt{{\nu}^2-k^2}$ due to the Eq.~(\ref{mass}), and 
due to the Eq.~(\ref{integrability}) the ${\nu}$
and $k$ are, correspondingly, the time- and the space-derivatives of
some field $u(t,x)$. With some abuse of language, one can say that
the $\tau$ upper half-plane $H_{+}$ plays the role of a target space 
in the effective theory.  

The lagrangian $L$~(\ref{infrared}) is invariant under the $S$-transformation,
\beq
L(\tau+1)=L(\tau)
\eeq
and nontrivially transforms under the $T$-transformation,
\beq
L(-\frac{1}{\tau})=L(\tau)+\frac{constant}{\tau}
\eeq
The $T$-transformation seems very plausible to be a duality transformation
between week and strong coupling regions. 

The dimensionality of the space-time doesn't
formally matter here, though, I doubt that such a complicated
lagrangian with no supersymmetry can give a renormalizable theory in arbitrary
dimension. However, in $(1+1)$-dimention almost any theory is renormalizable. 
From now on, I assume the $(1+1)$-dimensionality. In $(3+1)$-dimensional
theory, as long as one keeps at the tree level, this just
means a restriction of the kinematics considered. At quantum level the
$(1+1)$-dimensional kinematics might be necessary for the
renormalizability reason and can possibly be treated by two-step Whitham
averaging with two slow scales, the transversal being slower.

3. The effective theory obtained happens to be solvable at the classical
level. I am going now to construct a general solution of it. 
The Eq.~(\ref{integrability}) obviously solves as ${\nu}=u_{t}$,
$k=u_{x}$. Then the Legandre transformation
\beq 
\label{legandre}
{\psi}(\nu,k)={\nu}t+kx-u(t,x)
\eeq 
linearizes the Eq.~(\ref{kapitsa}), the resulting equation on $\psi$ being
\beq
\label{master}
((\partial_{\tau})^{2}
-\frac{\rho_{{\tau}{\tau}}}{\rho}(\partial_{\beta})^{2})\psi=0
\eeq 
where $\rho=\sqrt{{\nu}^{2}-k^{2}}$, $\beta$ is the rapidity parameter,
$th\beta=\frac{k}{\nu}$. Remind that $\rho$ is a function of $\tau$
via Eq.~(\ref{mass}). 

Remarkably, the $({\tau},{\beta})$-variables separate in Eq.~(\ref{master}),
and on the Fourier transform ${\psi}({\tau},{\xi})$ of
${\psi}({\tau},{\beta})$
\beq
\label{tarnsform}
\psi(\tau,\beta)=\int_{}^{}d{\xi}\psi(\tau,\xi)\exp(i{\beta}{\xi})
\eeq
there arises the equation
\beq
\label{masterm}
((\partial_{\tau})^{2}
+(\xi)^{2}\frac{\rho_{{\tau}{\tau}}}{\rho})\psi=0
\eeq    
What is much more remarkably, especially having in mind how 
complicated function of $\tau$  $\rho$ is (see Eq.~(\ref{mass}),
(\ref{pert})),
that the equation (\ref{masterm}) transforms to just a hypergeometric
equation. For that, notice that the equation~(\ref{masterm}) has nice
transformation properties.
Indeed, recall that the $g_{2}$ can be naturally considered as
an $SL(2,Z)$-invariant $2$-differential on the  ${\tau}$ upper half-plane.
The $\rho$ is then $SL(2,Z)$-invariant $(-\frac{1}{2})$-differential
and hence the potential term in Eq.~(\ref{masterm}) is a projective connection
and $\psi$ is another $SL(2,Z)$-invariant $(-\frac{1}{2})$-differential.
Therefore, it's sufficient to consider the Eq.~(\ref{masterm}) just in the
fundamental modular region $H_{+}/SL(2,Z)$. There is a global coordinate
$J$ in the fundamental region,
\beq
\label{coord}
J(\tau)=\frac{g_{2}^{3}(\tau)}{g_{2}^{3}(\tau)-27g_{3}^{2}(\tau)}
\eeq
in terms of which the equation (\ref{masterm}) becomes
\beq
\label{hyper}
((\partial_{J})^{2}+\frac{1-\delta_{1}^{2}}{4J^{2}}
+\frac{1-\delta_{2}^{2}}{4(J-1)^{2}}
+\frac{\delta_{1}^{2}+\delta_{2}^{2}-1}{4J(J-1)})\psi(J,\xi)=0
\eeq
where $\delta_{1}=\frac{1}{3}\sqrt{1-\frac{5\xi}{4}}$,
$\delta_{2}=\frac{1}{2}$. Important is that ${\psi}({\tau},{\xi})$ transforms
to $\psi(J,\xi)$ not as a function but as a $(-\frac{1}{2})$-differential. 
Apparently, the Eq.~(\ref{hyper}) solves in hypergeometric functions,
two linearly independent solutions, for example, being
\begin{eqnarray}
\label{basis}
\psi_{1}(J,\xi)=J^{\frac{1}{2}-\frac{1}{2}\delta_{1}}
(J-1)^{\frac{1}{4}}F(\frac{1}{4}-\frac{1}{2}\delta_{1},
\frac{1}{4}-\frac{1}{2}\delta_{1},
1-\delta_{1},J)\nonumber\\
\psi_{1}(J,\xi)=J^{\frac{1}{2}+\frac{1}{2}\delta_{1}}
(J-1)^{\frac{1}{4}}F(\frac{1}{4}+\frac{1}{2}\delta_{1},
\frac{1}{4}+\frac{1}{2}\delta_{1},
1+\delta_{1},J)
\end{eqnarray}
Notice that the parameters are such that one of two independent
solutions has logarithmic behaviour at $J=\infty$.
It's convenient to use another our $(-\frac{1}{2})$-differential, $\rho$,
to transform the above solutions from $J$ to $\tau$. $\rho(J)$ can be
obtained from the $\psi_{1}(J,\xi)$ at ${\xi}^{2}=-1$
\beq
\label{norm}
\rho=J^{\frac{1}{4}}(J-1)^{\frac{1}{4}}
\eeq   
The general solution of the problem Eq.~(\ref{master}) reads
\beq
\label{general}
\psi(\tau,\beta)=\rho(\tau)J^{-\frac{1}{4}}(J-1)^{-\frac{1}{4}}
\int_{}^{}d\xi[C_{1}(\xi)\psi_{1}(J,\xi)+C_{2}(\xi)\psi_{2}(J,\xi)]
\eeq
$\rho$ and J are now assumed to be the functions of $\tau$ as in
Eqs.~(\ref{mass}),(\ref{coord}).

The singularities in the Eq.~(\ref{hyper}) have the following
interpretation. $J={\infty}$ is the point about
which  the perturbative expansion goes. So the singularity at this point
can be identified as the 
perturbative Dyson's \cite{dyson} singularity. On the other hand side by the
duality transformation $\tau$ to $-\frac{1}{\tau}$ the perturbative
expansion is related
to the strong coupling expansion, and the singularity $J={\infty}$ can be 
equally viewed as the
strong coupling singularity (notice that the energy of the solution
Eq.~(\ref{solution}) at $J={\infty}$ is just the energy of the top of the
potential). 
The  $J=0$-singularity
corresponds to the $\phi$-particle becoming massless (see the comment in
brackets after the Eq.~(\ref{gparameter}). The $J=1$-singularity is the most
puzzling one. It corresponds to a self-dual
point under $\tau$ to $-\frac{1}{\tau}$ transformation. However, it's unknown
what are the dual describtions of the microscopic theory. Notice that even
without solving the equations,
the modular symmetry gives a great deal of information about the solution.
Say, very first term of the perturbative expansion of it immediately
gives information about the strong coupling expansion via the transformation
$\tau$ to $\frac{1}{\tau}$. 
Recall that the $J$-parameter or the $\tau$-parameter are, in fact, not
parameters, they are fields in the effective field theory (see comments
after formula~(\ref{infrared}). The lagrangian Eq.~(\ref{infrared}) in the
vicinity of the critical points becomes $L=({\nu}^{2}-k^{2})^{3}$ at $J=0$
and  $L=({\nu}^{2}-k^{2})^{\frac{1}{2}}$ at $J=1$. Curiously enough,
the equation corresponding to the latter lagrangian already appeared in
literature \cite{universal} and was christened ``universal''.

4.The effective theory can be used to describe the soft
multiparticle scattering. To do this in the tree approximation one needs
\cite{agks} a classical solution of the microscopic theory Eq.~(\ref{lagran})
obeying the asymptotic condition
\beq
\label{feynman}
\phi(t,x)=a(t,x){\exp}(-it)+a^{*}(t,x){\exp}(it)+O(\lambda)
\eeq
where $a(t,x)$, $a^{*}(t,x)$ are arbitrary slow functions of $(t,x)$
such that the exposed in Eq.~(\ref{feynman}) part of $\phi$ is
a solution of the free equation of motion (the equation of motion from
Eq.~(\ref{lagran}) without nonlinear term). 

The Fourier expansion of the Weierstrass function
\begin{eqnarray}
\label{fourier}
\wp(u|{\pi},{\pi}{\tau}')=-\frac{1}{12}+
\sum_{m=1}\frac{2mq^{2m}}{1-q^{2m}}\nonumber\\
-\sum_{m=1}
me^{-imu}\frac{q^{2m}}{1-q^{2m}}-
\sum_{m=1}me^{+imu}\frac{1}{1-q^{2m}}
\end{eqnarray}
($Im$ $u$ is assumed to be positive) 
helps rewriting the condition from the
Eq.~(\ref{feynman}) in terms of the effective theory entries:
\beq
\label{coupling}
q^{2}(1+O(q^{2}))=(\frac{\lambda}{6})^{2}aa^{*}
\eeq

\beq
\label{monodromy}
u(t,x)=t+\frac{1}{i}{\log}(q(\frac{a^{*}}{a})^{\frac{1}{2}})+O(q^{2})
\eeq
These conditions are supposed to fix the relevant solution of the 
Eq.~(\ref{kapitsa}). More detailed study of this subject will be
published elsewhere.

5.I believe that the method used here to describe the infrared behaviour
of some particular massive field theory model is in fact universal - like the
conformal field theory technique is universal in describing the infrared
behaviour in massless theories \cite{polyakov}. The main ingredient in the
describtion of this universality is modular geometry, the modular region
becoming a sort of target-space. For the
${\phi}^{3}$-theory the modular symmetry happens to be $SL(2,Z)$. Strictly
speaking, I can not garantie that it is not broken to some subgroup
(say, to $\Gamma(2)$ as in Seiberg-Witten case, or to other) by quantum 
effects.

The Seiberg-Witten effective theory
must be obtainable in a similar straightforward way -  one needs just to 
guess the relevant classical solution over which to develope the procedure
applied here to the ${\phi}^{3}$-model. The role of supersymmetry, so
important along the Witten-Seiberg ways, will just to garantie
renormalizability of the effective theory and a simple or
trivial scale dependence of its parameters. 
Such a construction would be very desirable since it would relate operators
in the microscopic theory and in the effective theory, which can for example
be used in obtaining $S$-matrix for $N=2$ $QCD$.
It would also
be very instructive to apply this technique to straightforwardly obtain
the S-matrixes, form-factors, etc in the models like the Gine-Gordon one.
And phenomenologists would just be happy with a similar construction
for realistic models, say, for the Standard one.  

As far as to the multiparticle scattering is concerned, the main lesson
is that it's not worth to discuss singularities (like the strong barion
number violation, etc.) before finding
a modular invariant (covariant) parametrization, in which all
the paradoxes - like the paradox with unitarization - are naturally resolved
and the strong coupling (large-number-of-particles) behaviour becomes
as cute as week coupling (small-namber-of-particles) one. Any perturbative
expansion - say, to compute quatum corrections to the tree amplitudes
- must respect that covariance to be reliable.

5. I'd like to thank A.Gorsky for collaboration on the initial stage
of the consideration \cite{agks} and V.Fock and A.Rosly for interesting
enlightening discussions. I appreciate I.Kogan's encouraging me to publish
this note. I am obliged to V.Fock and A.Rosly, V.Roubtsov, H.Braden,
I.Kogan, E.Corrigan and W.Zakrzewski for their really kind hospitality
in Bonn, in Paris,in Edinburgh, in Oxford and in Durham while the work
was done.  This work was partially supported by INTAS grant 93-633 and
by INTAS-CNRS grant 1010-CT-93-0023.

\end{document}